\documentclass{article}
\pdfoutput=1
\usepackage{waspaa17,amsmath,graphicx,url,times}
\usepackage{amsfonts}
\usepackage{color}
\usepackage{url}
\usepackage{hyperref}

\title{End-to-end Source Separation with Adaptive Front-Ends}

\twoauthors
  {Shrikant Venkataramani, Jonah Casebeer}
    {University of Illinois at Urbana Champaign \\
     svnktrm2, jonahmc2@illinois.edu}
  {Paris Smaragdis\thanks{This work was supported by NSF grant 1453104.}}
    {University of Illinois at Urbana Champaign \\
     Adobe Research}

\begin{document}

\ninept
\maketitle

\begin{sloppy}
\begin{abstract}

Source separation and other audio applications have traditionally relied on the use of short-time Fourier transforms as a front-end frequency domain representation step. The unavailability of a neural network equivalent to forward and inverse transforms hinders the implementation of end-to-end learning systems for these applications. We present an auto-encoder neural network that can act as an equivalent to short-time front-end transforms. We demonstrate the ability of the network to learn optimal, real-valued basis functions directly from the raw waveform of a signal and further show how it can be used as an adaptive front-end for supervised source separation. In terms of separation performance, these transforms significantly outperform their Fourier counterparts. Finally, we also propose a novel source to distortion ratio based cost function for end-to-end source separation.
\end{abstract}


\section{Introduction}
\label{sec:intro}
Several neural network~(NN) architectures and methods have been proposed for supervised single-channel source separation~\cite{smaragdis2017neural, huang2014deep, chandna2017monoaural, uhlich2017improving}. These approaches can be grouped together into a common two-step workflow as follows. The first step is to transform the time domain signals into a suitable time-frequency~(TF) representation using short-time Fourier transforms (STFTs). These short time spectra are subsequently divided into their magnitude and phase components. The actual separation procedure takes place in the second step of the workflow which often operates on the extracted magnitude components. Common approaches include neural networks which given the noisy magnitudes, either predict a noiseless magnitude spectrum~\cite{kim2015adaptive, grais2014deep, grais2017single}, or some type of a masking function \cite{zhang2016deep}. Figure~\ref{fig:block_diagram} (a) shows the block diagram of such a system using the STFT as a front-end transform.
\begin{figure}[ht]
\centering
  \includegraphics[clip, trim = 0cm 0cm 0cm 0cm, width=\linewidth]{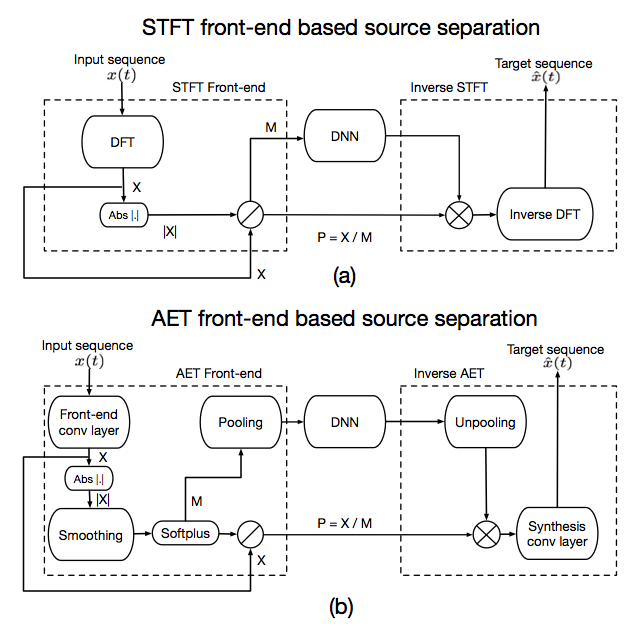}
  \caption{Block diagram of generalized NN based source separation system using (a) STFT front-end (top) and (b) the proposed adaptive front-end transform (bottom).}~\label{fig:block_diagram}
\end{figure}
Although they produce very good results, these NN approaches suffer from a couple of drawbacks. First, by restricting the processing to only magnitudes, they do not take full advantage of the information contained in the input signals. Additionally, there is no guarantee that the STFT (or whichever front-end one uses) would be optimal for the task at hand.
In this paper, we investigate the use of adaptive front-end transforms for supervised source separation. Using these adaptive front-ends for forward and inverse transformations enables the development of end-to-end learning systems for supervised source separation and potentially for other related NN models that rely on fixed transforms. In section~\ref{sec:STCT}, we consider the use of the DCT as a front-end transform to develop the necessary intuition for using real-valued front-ends. We then develop a neural network equivalent in section~\ref{sec:AutoencoderTransformsForSourceSeparation} and show how it can be used as an adaptive front-end for end-to-end source separation. Our experiments and results are discussed in section~\ref{sec:Experiments} and we conclude in section~\ref{sec:Conclusion}.
\begin{figure}[t]
\centering
  \includegraphics[clip, trim = 4cm 0.4cm 6.5cm 0.35cm, width=0.95\columnwidth]{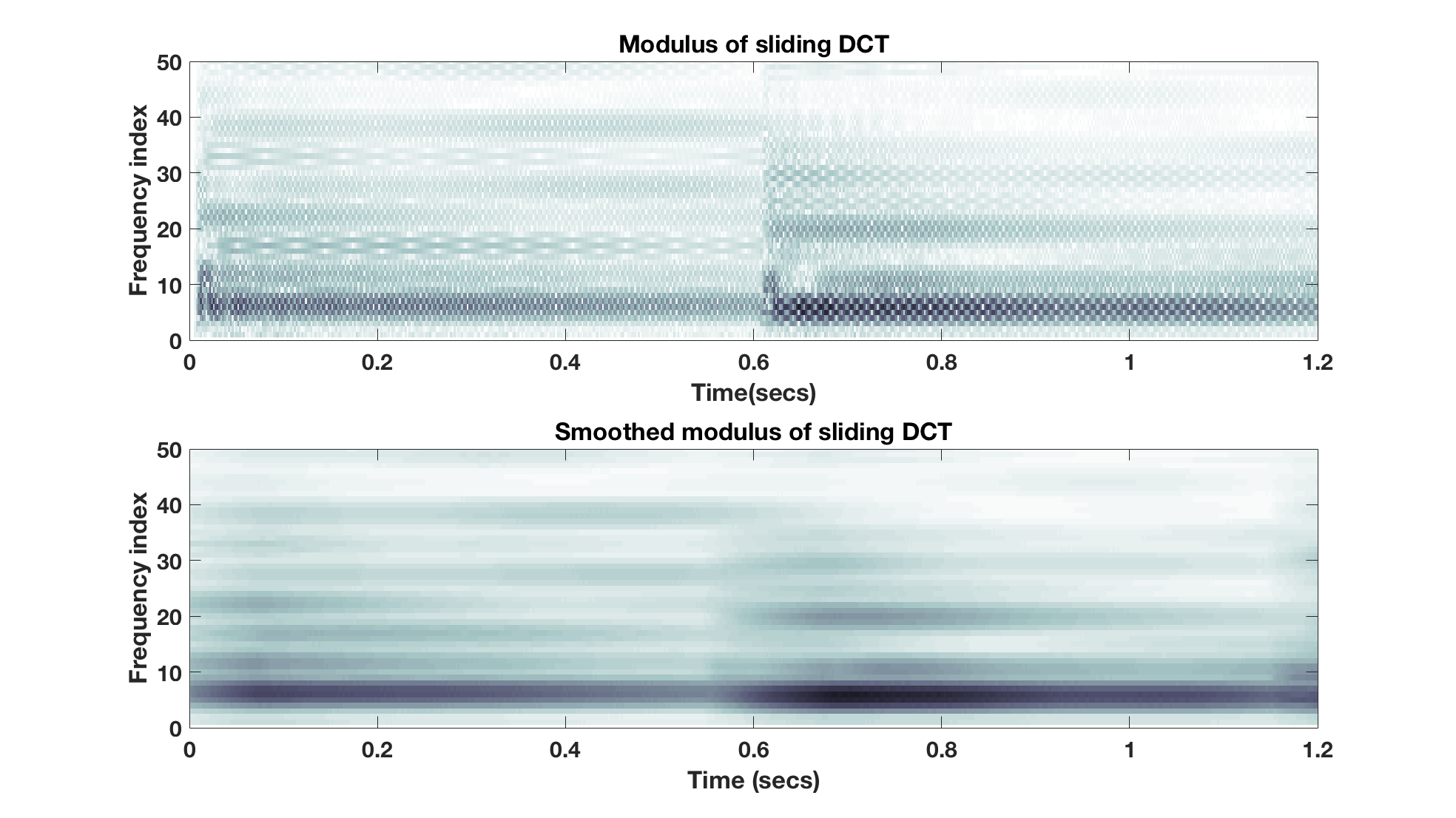}
  \caption{Modulus and smoothed coefficients of a sliding DCT transform (first $10$ coefficients) for a sequence of piano notes. The unsmoothed coefficients oscillate excessively and need to be averaged across time to produce what we would expect as a magnitude spectrum}~\label{fig:dct_interplay}
\end{figure}
%
\section{Using Real-valued Transforms}
\label{sec:STCT}
Given that ultimately we want to replace the front-end transform with a regular convolutional layer, we will first use a fixed real-valued transform instead of the DFT. This will allow us to develop an appropriate formulation before we move to an adaptive transform. Given a time domain sequence~$x$, the short-time transform operation of~$x$ can be expressed as a generalized equation given by,
\begin{equation}
    \mathbf{X}_{nk} = \sum_{t=0}^{N-1} x(nh+t) \cdot w(t) \cdot b(k,t)
\label{eq:short_time_transform}
\end{equation}
Here, $\mathbf{X}_{nk}$ represents the coefficient corresponding to the $k^{\text{th}}$ component in the $n^{\text{th}}$ frame, $N$ represents the size of a window function~$w$ and $h$ represents the hop size of the short-time transform. The functions $b(k,t)$ form the basis functions of the transformation.
%
%
For illustration, we will use the type-2 DCT basis functions for the values of $b(k,t)$. As shown in figure~\ref{fig:dct_interplay}, the resulting spectral energies do not maintain locality~\cite{abdel2014convolutional} and exhibit a modulation that could be dependent on the frequency, the window size and the hop size parameters. Thus, we need to apply a suitable averaging (smoothing) operation across time. This can be achieved by convolving the absolute values of the resulting coefficients with a suitable averaging function across the time axis given as,
\begin{equation}
    \mathbf{M} = \left|\mathbf{X}\right|*s~
\label{eq:smoothing_stage}
\end{equation}
Here, $|\cdot|$ represents the element-wise modulus operator, $s$ represents the bank of smoothing filters and $*$ denotes the one-dimensional convolution operation applied only along the time axis. The matrix~$\mathbf{M}$ obtained after smoothing can be interpreted as the magnitude spectrogram of the signal in this representation. The variations in the coefficients that are not captured by this magnitude spectrogram~$\mathbf{M}$ can be captured in a new matrix given by $\mathbf{P} = \frac{\mathbf{X}}{\mathbf{M}}$ where, the division is also element-wise. This can be interpreted as the corresponding phase component of the sequence $x$. We can also interpret these two quantities as the results of a demodulation operation in which we extract a smooth amplitude modulator, which modulates by a faster moving carrier that encodes more of the details. Using this approach we can easily match the performance of the STFT front-end, while only performing real-valued computations.
However, the use of a fixed front-end transform continues to pose some challenges. Short-time transforms need to be optimized with respect to the window size, window shape and hop size parameters. In the case of real transforms, the transformation must be followed by a smoothing operation that depends on the window size, the hop size and coefficient frequency. Thus, we also need to optimize over suitable smoothing function shapes and durations. As described in section~\ref{sec:AutoencoderTransformsForSourceSeparation}, we can interpret each step of the forward short-time transform as a neural network. In doing so, we can automatically learn adaptive transforms and smoothing functions directly from the waveform of the audio signal, thereby bypassing the aforementioned issues.
\begin{figure*}[ht!]
\centering
  \includegraphics[clip, trim = 0cm 0cm 0cm 0cm, width = \textwidth]{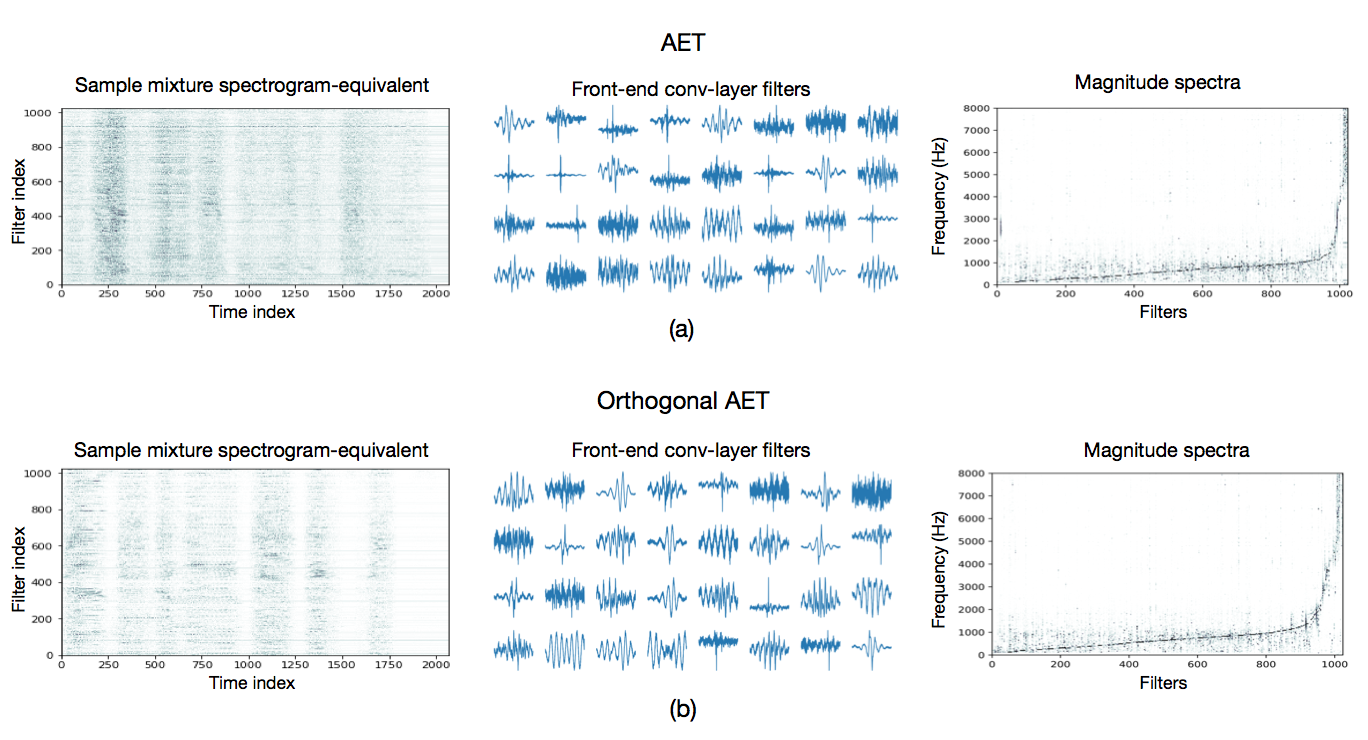}
  \caption{(a) A sample mixture spectrogram, front-end convolutional layer filters, and their normalized magnitude spectra using the AET (top) (b) A sample mixture spectrogram, front-end convolutional layer filters, and their normalized magnitude spectra using the orthogonal-AET (bottom). The orthogonal-AET uses a transposed version of the analysis filters for the synthesis convolutional layer. The filters are ordered according to their dominant frequency component (from low to high). In the middle subplots, we show a subset of the first $32$ filters.}~\label{fig:AETbases}
\end{figure*}
%
\section{Auto-encoder Transforms for Source separation}
\label{sec:AutoencoderTransformsForSourceSeparation}
Despite recent advances in neural networks, the STFT continues to be the transform of choice for audio applications~\cite{smaragdis2017neural, huang2014deep, chandna2017monoaural, uhlich2017improving}. Recently, Sainath et.al~\cite{sainath2015learning}, and Dieleman and Schrauwen~\cite{dieleman2014end} have proposed the use of a convolutional layer as an alternative to front-end STFTs. In this section, we expand upon the premise and develop a real valued, convolutional auto-encoder transform~(AET) that can be used as an alternative to front-end short-time transforms. The encoder part of the auto-encoder~(AE) acts as the analysis filter bank and produces a spectrogram-equivalent of the input. The decoder performs the synthesis step used to recover the time-domain signal.
\subsection{Analysis Encoder}
\label{subsec:AnalysisEncoder}
Assuming a unit sample hop size, we can interpret~(\ref{eq:short_time_transform}) as a filtering operation,
\begin{equation}
    \mathbf{X}_{nk} = \sum_{t=0}^{N-1} x(n+t) \cdot \mathbf{F}(k,t)~
\label{eq:short_time_transform_as_cnn}
\end{equation}
Thus, we may replace the front-end transform by a convolutional neural network~(CNN) such that the $k^{\text{th}}$ filter of the CNN represents the $k^{\text{th}}$ row of $\mathbf{F}$. The output of the CNN gives a feature space representation of the input signal with a unit hop size.
As described in section~\ref{sec:STCT}, the transformation stage should be followed by a smoothing operation. This smoothing stage given by~(\ref{eq:smoothing_stage}) can also be interpreted as a CNN applied on $\left|\mathbf{X}\right|$. However, since there are no non-negativity constraints applied on the averaging filters, the elements of the smoothed spectrogram~$\mathbf{M}$ can potentially assume negative values. To avoid this solution, we include a non-linearity to the convolutional smoothing layer. The non-linearity $g: \mathbb{R} \rightarrow \mathbb{R}^{+}$ is a mapping from the space of real numbers to the space of positive real numbers. In this paper, we use a softplus non-linearity for this step. As before, the output of this layer~$\mathbf{M}$ can be interpreted as the magnitude spectrogram of the input signal and $\mathbf{P} = \frac{\mathbf{X}}{\mathbf{M}}$ can be interpreted as the corresponding phase. The phase component captures the high-frequency variations in the coefficients which cannot be modeled in the smoothed spectrogram. In order to use a more economical subsampled representation, we can apply a max-pooling layer \cite{springenberg2014striving} that replaces a pool of $h$ frames with a single frame containing the maximum value of each coefficient over the pool. Note that all the convolution and pooling operations are one-dimensional i.e., the operations are applied across the time-axis only. In addition, these operations are independently applied on each filter output of the front-end CNN.
\subsection{Synthesis Decoder}
\label{subsec:SynthesisDecoder}
Given the magnitude and phase equivalents obtained using the AET, the next step is to synthesize the signal back into the time domain. This can be achieved by inverting the operations performed by the analysis encoder while computing the forward transform.
The first step of the synthesis procedure is to undo the effect of the lossy pooling layer. We use an upsampling operator by inserting as many zeros between the frames as the pooling duration as proposed by Zieler et.al., ~\cite{zeiler2014visualizing}. The unpooled magnitude spectrogram is then multiplied by the phase using an element-wise multiplication to give an approximation $\mathbf{\hat{X}}$ to the matrix~$\mathbf{X}$. We invert the operation of the first transform layer by a convolutional layer to implement the deconvolution operation. This convolutional layer thus, performs the interpolation between the samples. Essentially, the output of the analysis encoder gives the weights of each basis function in representing the time domain signal. The synthesis layer reconstructs each component by adding multiple shifted versions of the basis functions at appropriate locations in time. This inversion procedure is similar to the filterbank-summation technique of inverting the subsampled STFT representation of a sequence~\cite{smith2011spectral}. The weights (filters) of the first convolutional layer give the AET basis functions (see figure~\ref{fig:AETbases}).
%
\subsection{Examining the learned bases}
\label{subsec:PracticalConsiderations}
Having developed the convolutional auto-encoder for AETs, we can now examine and understand the nature of the basis functions obtained. We plot the bases and their corresponding normalized magnitude spectra in figure~(\ref{fig:AETbases}) for AET (top) and orthogonal-AET (bottom) transforms. In the case of orthogonal-AET, the synthesis convolutional layer filters are held to be transposed versions of the front-end layer filters. Thus, the inverse transform is a transpose of the forward transform. In figure~\ref{fig:AETbases}, the middle figures are the filters obtained from the front-end convolutional layer that operates on the input mixture waveform. The complete network architecture and training-data for obtaining these plots are described in section~\ref{subsec:ExperimentalSetup}. We rank the filters according to their dominant frequency component. Then, we use a $1024$-point Fourier transform to compute the magnitude spectra of the filters. The middle figures show the first $32$ low-frequency filters obtained after the sorting step. The plots on the right show the corresponding filter magnitude spectra. From the magnitude spectra, it is clear that the filters are frequency selective even though they are noisy and consist of multiple frequencies. The filters are concentrated at the lower frequencies and spaced out at the higher frequencies. The left figures show the output of the front-end layer for a sample mixture input waveform, with respect to the corresponding transform bases. These observations hold for the AET and the orthogonal-AET. In other words, we see that the adaptive front-ends learn a representation that is tailored to the input waveform.

\subsection{End-to-end Source Separation}
\label{subsec:End-to-endSourceSeparation}
Figure~\ref{fig:block_diagram}(b) shows the application of AET analysis and synthesis layers for end-to-end supervised source separation. The forward and inverse transforms can be directly replaced by the AET analysis and synthesis networks in a straightforward manner. We train the network by giving the mixture waveform at the input and minimize a suitable distance metric between the network output and the clean waveform of the source. Thus, the network learns to estimate the contribution of the source given the raw waveform of the mixture. Since the basis and smoothing functions are automatically learned, it is reasonable to expect the network to learn optimal, task specific basis and smoothing functions.
The advantages of interpreting the forward and inverse transforms as a neural network now begins to come through. We can propose interesting extensions to these adaptive front-ends by exploiting the available diversity of neural networks. For example, we can propose recurrent auto-encoder alternatives as in~\cite{shrikant2017neuralnetworkalternatives} or multilayer or recurrent extensions to adaptive front-ends. We can also experiment with better pooling strategies for the adaptive front-end. These generalizations to adaptive front-ends are not easily explored with fixed front-end transforms.
%
\section{Experiments}
\label{sec:Experiments}
We now present some experiments aimed at comparing the effect of different front-ends for a supervised source separation task. We evaluate the separation performance for three types of front-ends: STFT, AET and orthogonal-AET. We compare the results based on the BSS\_EVAL metrics.~\cite{fevotte2005bss_eval}.
\subsection{Experimental setup}
\label{subsec:ExperimentalSetup}
For training the neural networks, we randomly selected $10$ male-female speaker pairs from the TIMIT database~\cite{timit}. In the database, each speaker has a total of $10$ recorded sentences. For each pair, we mix the sentences at $0$ dB. This gives $10$ mixture sentences per pair and a total of $100$ mixture sentences overall. For each pair, we train on $8$ sentences and test on the remaining two. Thus, the network is trained on $80$ mixture sentences and evaluated on the remaining $20$ mixture sentences. For training the NN we use a batch size of $16$ and a dropout-rate of $0.2$.
The separation network consisted of a cascade of $3$ dense layers with $512$ hidden units each, each followed by a softplus non-linearity, which is the architecture used in ~\cite{kim2015adaptive}. As seen in figure~\ref{fig:block_diagram}, the STFT and AET magnitude spectrograms were given as inputs to the separator network. The STFT was computed at a window-size of $1024$ samples at a hop of $16$ samples. For the STFT front-end, the separation results were inverted into the time domain using the inverse STFT operation and the mixture phase. To have a fair comparison with adaptive front-ends, the CNN filters were chosen to have a width of $1024$, a stride of $16$ samples was selected for the pooling operation. The smoothing layer was selected to have a length of $5$ frames.
We used two different cost functions for this task. First we used the mean-squared error (MSE) between the target waveform and the target estimate. Second, we used a cost function that directly optimizes the signal to distortion ratio (SDR) instead. To do the latter, we note that for a reference signal $y$ and an output $x$ we would maximize:
\begin{multline}
\label{eq:sdrcost}
    \text{max}~\text{SDR}(x,y) = \text{max}\frac{\left<xy\right>^2}{\left<yy\right>~\left<xx\right> - \left<xy\right>^2} \equiv \\ \text{min}\frac{\left<yy\right>\left<xx\right> - \left<xy\right>^2}{\left<xy\right>^2} = \text{min}\frac{\left<yy\right>\left<xx\right>}{\left<xy\right>^2} - \frac{\left<xy\right>^2}{\left<xy\right>^2} \propto \text{min}\frac{\left<xx\right>}{\left<xy\right>^2}
\end{multline}
Thus, maximizing the SDR is equivalent to maximizing the correlation between $x$ and $y$, while producing a minimum energy output.
%
%
\begin{figure}[t]
\centering
  \includegraphics[clip, trim = 8cm 3cm 9cm 3cm, width=0.98\columnwidth]{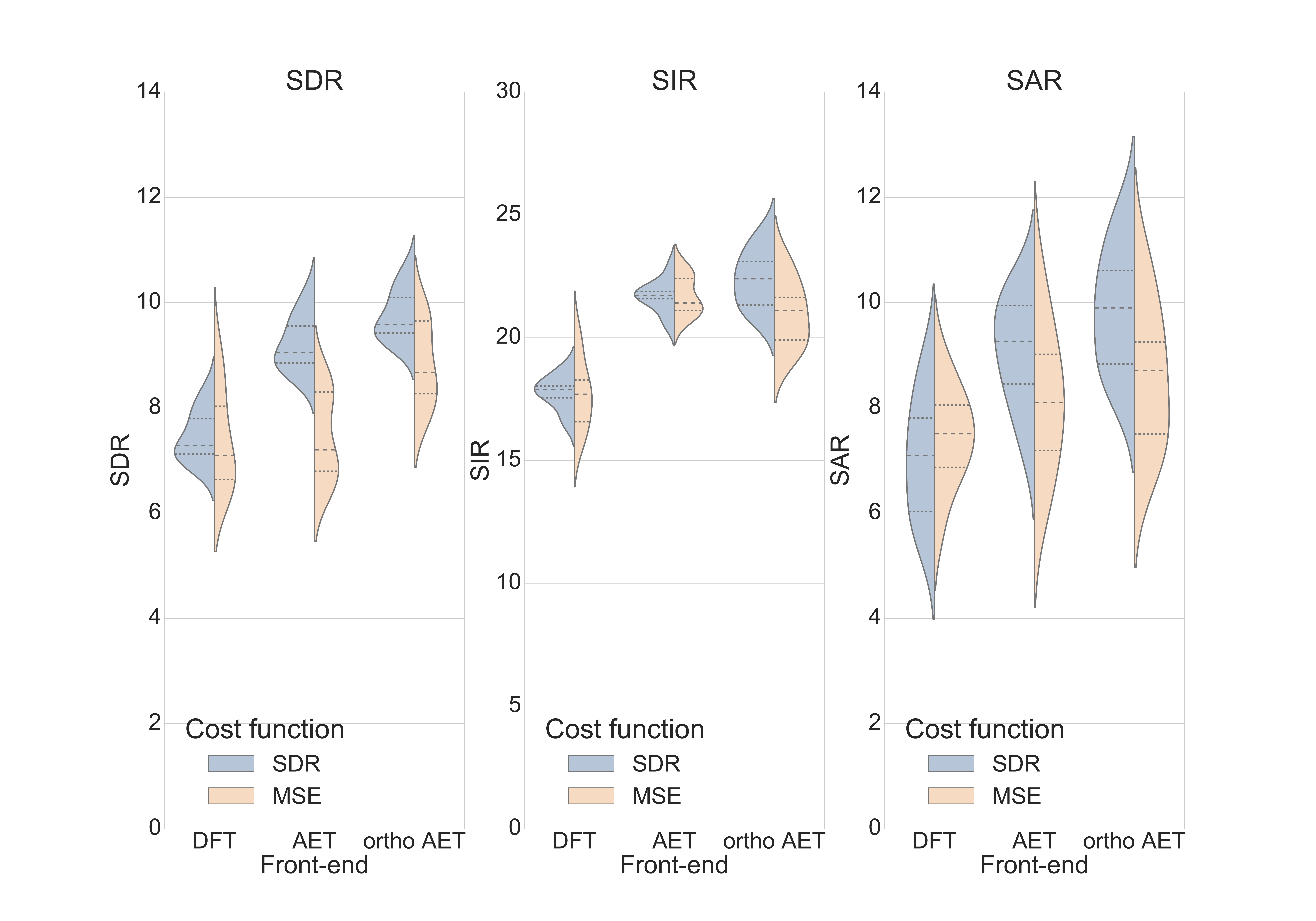}
  \caption{Comparison of source separation performance on $20$ speech on speech mixtures in terms of BSS\_EVAL parameters. We compare the separation performance for multiple front end transforms viz., STFT, AET and orthogonal AET. The dashed line in the centre indicates the median value and the dotted lines above and below indicate the interquartile range. We see that opting for an adaptive front-end results in a significant improvement in source separation performance over STFT front-ends. Comparing the cost-functions we see that SDR (left) is a more appropriate cost-function to MSE (right) for end-to-end source separation.}~\label{fig:SourceSeparationSpeech}
\end{figure}
\subsection{Results and Discussion}
\label{subsec:ResultsAndDiscussion}
The corresponding violin plots that show the distribution of the BSS\_EVAL metrics from our experiments are shown in figure~\ref{fig:SourceSeparationSpeech}. We see that the use of AETs improves the separation performance in terms of of all metrics compared to an STFT front-end. We additionally see that when using the orthogonal AET we obtain additional performance gains, overall in the range of $2$dB for SDR, $5$dB in SIR and $3$dB in SAR. One possible reason for the increased performance of the orthogonal AET could be the reduction in the number of trainable parameters caused by forcing the synthesis transform to be the transpose of the analysis transform, which in turn reduces the possibility of over-fitting to the training data. The above trends appear consistent for both the cost-functions considered.
Additionally we can compare the use of the two cost-functions for our networks. We see that the use of SDR as a cost function (expectedly) results in a significant improvement over using MSE. This is observed for all the front-end options considered in this paper. We also note that the use of MSE increases the variance of the separation results, whereas the SDR is more consistent. We thus conclude that the SDR is a better choice of cost-function for end-to-end source separation.

\section{Conclusion and Future Work}
\label{sec:Conclusion}
In this paper, we developed and investigated a convolutional auto-encoder based front-end transform that can be used as an alternative to using STFT front-ends. The adaptive front-end comprises a cascade of three layers viz., a convolutional front-end transform layer, a convolutional smoothing layer and a pooling layer. We have shown that AETs are capable of automatically learning adaptive basis functions and discovering data-specific frequency domain representations directly from the raw waveform of the data. The use of AETs significantly improves the separation performance over fixed front-ends and also enables the development of end-to-end source separation. Finally, we have also demonstrated that the SDR is a superior alternative as a cost-function for end-to-end source separation systems.
%
%

\bibliographystyle{IEEEtran}
\bibliography{refs17}
%
%
%
%
%
%
%
%
%

\end{sloppy}
\end{document}